\documentclass[sigconf]{aamas}  

\AtBeginDocument{%
  \providecommand\BibTeX{{%
    \normalfont B\kern-0.5em{\scshape i\kern-0.25em b}\kern-0.8em\TeX}}}
    
\usepackage{booktabs}    
\usepackage{flushend} 

\setcopyright{ifaamas}  
\copyrightyear{2020} 
\acmYear{2020} 
\acmDOI{} 
\acmPrice{} 
\acmISBN{} 
\acmConference[AAMAS'20]{Proc.\@ of the 19th International Conference on Autonomous Agents and Multiagent Systems (AAMAS 2020)}{May 9--13, 2020}{Auckland, New Zealand}{B.~An, N.~Yorke-Smith, A.~El~Fallah~Seghrouchni, G.~Sukthankar (eds.)}  

\usepackage{amsmath,amsfonts,bm}

\def\eqref#1{equation~\ref{#1}}

\def\1{\bm{1}}

\DeclareMathAlphabet{\mathsfit}{\encodingdefault}{\sfdefault}{m}{sl}
\SetMathAlphabet{\mathsfit}{bold}{\encodingdefault}{\sfdefault}{bx}{n}

\usepackage{booktabs}
\usepackage{blkarray}
\usepackage{hyperref}
\usepackage{subfigure}
\usepackage{times}  
\usepackage{helvet} 
\usepackage{courier}  
\usepackage{graphicx} 
\usepackage{amsmath}
\usepackage{amsfonts}
\usepackage{amsmath,amsthm}
\usepackage{amsfonts}
\usepackage{graphicx}  
\usepackage{caption}
\usepackage{multirow}
\usepackage{wrapfig}
\usepackage{wrapfig}
\usepackage{xfrac}
\usepackage{algorithm}
\usepackage[noend]{algpseudocode}

\usepackage[most]{tcolorbox}

\expandafter\def\expandafter\normalsize\expandafter{%
    \normalsize
    \setlength\abovedisplayskip{6pt}
    \setlength\belowdisplayskip{6pt}
    \setlength\abovedisplayshortskip{6pt}
    \setlength\belowdisplayshortskip{6pt}
}
\usepackage{flushend}

\begin{document}

\title{$\alpha^{\alpha}$-Rank: Practically Scaling $\alpha$-Rank  through \\Stochastic Optimisation}  


\author{Yaodong Yang}
\authornote{Equal contribution.}
\affiliation{%
  \institution{Huawei Technologies R\&D U.K.}
    \institution{University College London}
}
\email{yaodong.yang@huawei.com}
\author{Rasul Tutunov$^*$}
\affiliation{%
  \institution{Huawei Technologies R\&D U.K.}
}
\email{rasul.tutunov@huawei.com}

\author{Phu Sakulwongtana}
\affiliation{%
  \institution{Huawei Technologies R\&D U.K.}
}
\email{phu.sakulwongtana@huawei.com}

\author{Haitham Bou Ammar}
\affiliation{%
  \institution{Huawei Technologies R\&D U.K.}
      \institution{University College London}
}
\email{haitham.ammar@huawei.com}

\begin{abstract}
Recently, $\alpha$-Rank, a graph-based algorithm, has been proposed as a solution to ranking joint policy profiles in large scale multi-agent systems. $\alpha$-Rank claimed tractability through a polynomial time implementation with respect to the total number of pure strategy profiles. Here, we note that inputs to the algorithm were not clearly specified in the original presentation; as such, we deem complexity claims as not grounded, and conjecture solving $\alpha$-Rank is NP-hard. 

The authors of $\alpha$-Rank suggested that the input to $\alpha$-Rank can be an exponentially-sized payoff matrix; a claim promised to be clarified in subsequent manuscripts. Even though $\alpha$-Rank exhibits a polynomial-time solution with respect to such an input, we further reflect additional critical problems. We demonstrate that due to the need of constructing an exponentially large Markov chain, $\alpha$-Rank is infeasible beyond a small finite number of agents. We ground these claims by adopting amount of dollars spent as a non-refutable evaluation metric. Realising such scalability issue, we present a stochastic implementation of $\alpha$-Rank with a double oracle mechanism allowing for reductions in joint strategy spaces. Our method, $\alpha^\alpha$-Rank, does not need to save exponentially-large transition matrix, and can terminate early under required precision. Although theoretically our method exhibits similar worst-case complexity guarantees compared to $\alpha$-Rank, it allows us, for the first time, to practically conduct large-scale multi-agent evaluations. On $10^4 \times 10^4$ random matrices, we achieve $1000x$ speed reduction. Furthermore, we also show successful results on large joint strategy profiles with a maximum size in the order of  $\mathcal{O}(2^{25})$ ($\approx 33$ million joint strategies) -- a setting not evaluable using $\alpha$-Rank with reasonable computational budget.

\end{abstract}


\maketitle

\section{Introduction}

Scalable policy evaluation and learning have been long-standing challenges in multi-agent reinforcement learning (MARL) with two difficulties obstructing progress. 
First, joint-strategy spaces exponentially explode when a large number of strategic decision-makers is considered, and second, the underlying game dynamics may exhibit cyclic behaviour (e.g. the game of Rock-Paper-Scissor) rendering an appropriate evaluation criteria non-trivial.
Focusing on the second challenge, much work in multi-agent systems followed a game-theoretic treatment proposing fixed-points, e.g., Nash \cite{nash1950equilibrium} equilibrium, as potentially valid evaluation metrics \cite{yang2018mean,zhang2019bi}. 
Though appealing, such measures are normative only when prescribing behaviours of perfectly rational agents -- an assumption rarely met in reality \cite{grau2018balancing,wen2019probabilistic, Felix, wen2019multi}. 
In fact, many game dynamics have been proven not converge to any fixed-point equilibria \cite{hart2003uncoupled,viossat2007replicator,yang2018study}, but rather to limit cycles  
\cite{palaiopanos2017multiplicative,bowling2001convergence}. 
Apart from these challenges, solving for a Nash equilibrium even for ``simple'' settings, e.g. two-player games is known to be PPAD-complete \cite{chen2005settling} -- a demanding complexity class when it comes to computational  requirements.

To address some of the above limitations,  \cite{omidshafiei2019alpha} recently proposed $\alpha$-Rank as a graph-based game-theoretic solution to multi-agent evaluation. 
$\alpha$-Rank adopts Markov Conley Chains to highlight the presence of cycles in game dynamics, and attempts to compute stationary distributions as a mean for strategy profile ranking. 
In a novel attempt, the authors reduce multi-agent evaluation to computing a stationary distribution of a Markov chain. Namely, consider a set of $N$ agents each having a strategy pool of size $k$, a Markov chain is, first, defined over the graph of joint strategy profiles with a transition matrix $\bm{T} \in \mathbb{R}^{k^{N} \times k^{N}}$, and then a stationary distribution $\bm{\nu} \in \mathbb{R}^{k^{N}}$ is computed solving: $\bm{T}\bm{\nu} = \bm{\nu}$. 
The probability mass in $\bm{\nu}$ then represents the ranking of joint-strategy profile. 

Extensions of  $\alpha$-Rank have been developed on various instances.
\cite{rowl2019multiagent} adapted $\alpha$-Rank to model games with incomplete information. 
\cite{muller2019generalized} combined $\alpha$-Rank with the policy search space oracle (PSRO) \cite{lanctot2017unified} and claimed their method to be a \emph{generalised} training approach for multi-agent learning. 
Unsurprisingly, these work inherit the same claim of tractability from $\alpha$-Rank. 
For example, the abstract in~\cite{muller2019generalized} reads ``$\alpha$-Rank, which
is unique (thus faces no equilibrium selection issues, unlike Nash) and \emph{tractable}
to compute in general-sum, \emph{many-player} settings."

In this work, we contribute to refine the claims made in $\alpha$-Rank dependent on its input type. We thoroughly argue that $\alpha$-Rank exhibits a prohibitive computational and memory bottleneck that is hard to remedy even if pay-off matrices were provided as inputs. We measure such a restriction using money spent as a non-refutable metric to assess $\alpha$-Rank's validity scale. With this in mind, we then present a stochastic solver that we title $\alpha^{\alpha}$-Rank as a scalable and memory efficient alternative. Our method reduces memory constraints, and makes use of the oracle mechanism for reductions in joint strategy spaces. This, in turn, allows us to run large-scale multi-player experiments, including evaluation on self-driving cars and Ising models where the maximum size involves tens of millions of joint strategies.  
\section{A Review of $\alpha$-Rank} \label{Sec:MC}
\begin{figure*}[t!]
    \centering
        \vspace{-25pt}
    \subfigure{\label{fig:a}\includegraphics[width=.9\textwidth]{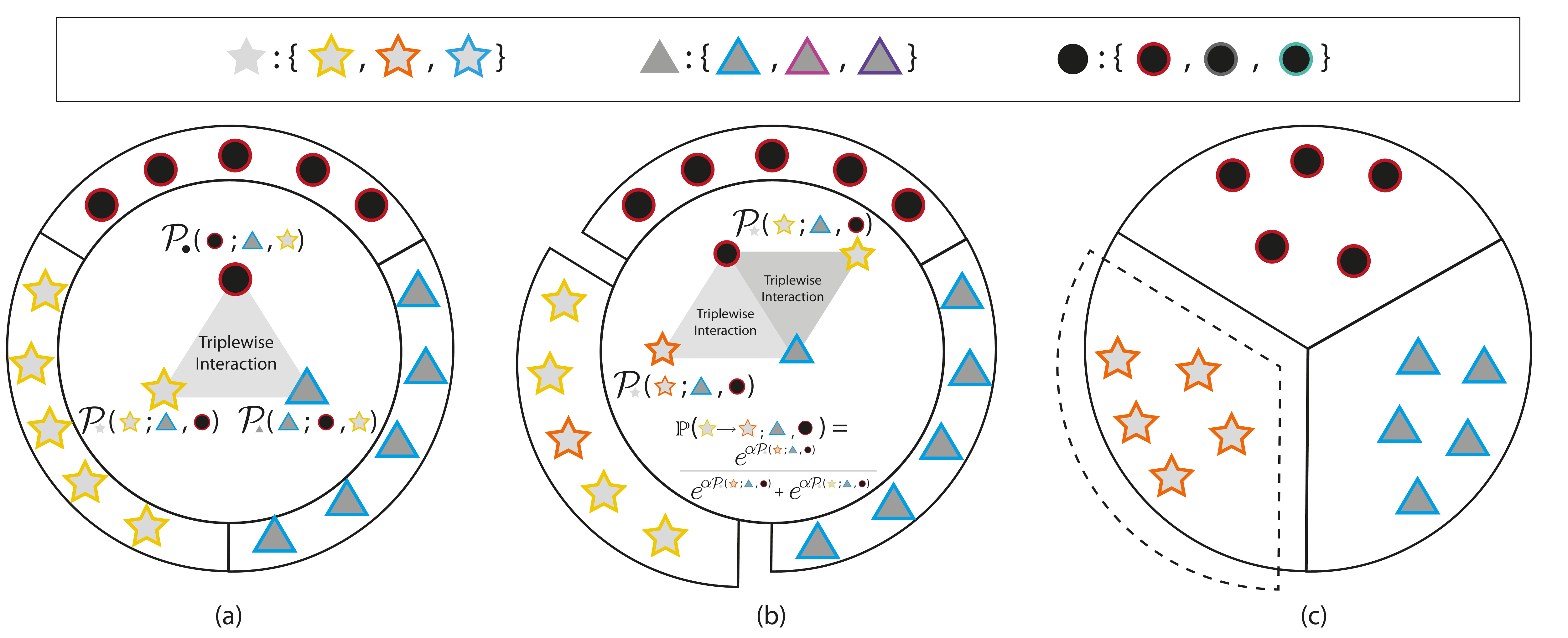}}
            \vspace{-15pt}
    \caption{Example of population based evaluation on $N=3$ players (star, triangle, circle) each with $|s| = 3$ strategies (denoted by the colours) and $m=5$ copies. a) Each population obtains a fitness value $\mathcal{P}_i$ depending on the strategies chosen, b) one  mutation strategy (red star) occurs, and c) the population either selects the original strategy, or being fixated by  the mutation strategy.}\label{fig:nfgH}
        \vspace{-10pt}
\end{figure*}
In $\alpha$-Rank, strategy profiles of $N$ agents are evaluated through an evolutionary process of mutation and selection. Initially, agent populations are constructed by creating multiple copies of each learner $i \in \{1, \dots, N\}$ assuming that all agents (in one population) execute the same unified policy. 
With this, $\alpha$-Rank then simulates a multi-agent game played by randomly sampled learners from each population. Upon game termination, each participating agent receives a payoff to be used in policy mutation and selection after its return to the population. 
Here, the agent is faced with a probabilistic choice between switching to the mutation policy, continuing to follow its current policy, or randomly selecting a novel policy (other than the previous two) from the pool. 
This process repeats with the goal of determining an evolutionary dominant profile that spreads across the population of agents. 
Fig.~\ref{fig:nfgH} demonstrates a simple example of a three-player game, each playing three strategies. 

\emph{Mathematical Formulation:} To formalise $\alpha$-Rank, we consider $N$ agents each, denoted by $i$, having access to a set of strategies of size $k_{i}$. 
We refer to the strategy set for agent $i$ by $\mathcal{S} _{i} = \big\{ \pi_{i,1} , \dots, \pi_{i, k_{i}} \big\}$, $k_i=|\mathcal{S}_i|$, with $\pi_{i, j}  : \mathcal{X} \times \mathcal{A}_{i} \rightarrow [0,1]$ representing the $j^{th}$ allowed policy of the learner. 
$\mathcal{X}$ represents the set of states and $\mathcal{A}_{i}$ is the set of actions for agent $i$.
A joint strategy profile is a set of  policies for all participating agents in the joint strategy set, i.e., $\mathcal{S}_{\text{joint}} = \mathcal{S} _{1} \times \mathcal{S} _{2} \times \dots \times \mathcal{S} _{N}$: $\pi_{\text{joint}}  = \big\{ \pi_{1, j_{1}},...,\pi_{N, j_{N}} \big\}$, with $\pi_{i, j_{i}} \in \mathcal{S}_{i} $ and $j_{i} \in \{1, \dots, k_{i}\}$. We assume $k=k_1 =\ldots = k_N$ hereafter.

To evaluate performance, we assume each agent is additionally equipped with a payoff (reward) function $\mathcal{P}_{i} : \mathcal{S}_{\text{joint}}\rightarrow \mathbb{R}_{+}$. Crucially, the domain of $\mathcal{P}_{i} $ is the pool of \emph{joint strategies} so as to accommodate the effect of other learners on the $i^{th}$ player's performance. Finally, given a joint profile $\pi_{\text{joint}} $, we define the corresponding joint payoff to be the collection of all individual payoff functions, i.e., $\mathcal{P}_{\text{joint}}  =\big\{\mathcal{P}_{1} \big(\pi_{\text{joint}} \big), \dots, \mathcal{P}_{N} \big(\pi_{\text{joint}} \big)\big\}$. 
After attaining payoffs from the environment, each agent returns to its population and faces a choice between switching the whole population to a mutation policy, exploring a novel policy, or sticking to the current one. 
Such a choice is probabilistic and defined  proportional to rewards by 
\begin{align*}
\mathbb{P}(\pi _{i,a}\to \pi _{i,b},\boldsymbol{\pi} _{-i}) &= \frac{e^{\alpha \mathcal{P} _i(\pi _{i,b},\boldsymbol{\pi} _{-i})}}{e^{\alpha \mathcal{P} _i(\pi _{i,a},\boldsymbol{\pi} _{-i})} + e^{\alpha \mathcal{P} _i(\pi _{i,b},\boldsymbol{\pi} _{-i})}} - \frac{\mu}{2} \\  & \ \ \ \ \ \ \ \ \ \ \ \ \ \ \ \ \ \  \ \ \text{for $\left(\pi_{i,a} , \pi_{i,b} \right) \in \mathcal{S}_{i}  \times \mathcal{S}_{i} $,} \\
\mathbb{P}(\pi _{i,a}\to \pi _{i,c},\boldsymbol{\pi} _{-i}) &= \frac{\mu}{K_i - 2},\ \ \ \ \forall \pi _{i,c}\in\mathcal{S} _i\setminus \{\pi _{i,a}, \pi _{i,b}\},
\end{align*}
with $\mu \in \mathbb{R}_{+}$ being an exploration parameter\footnote{In $\alpha$-Rank, $\mu$ is heuristically set to a small positive constant to ensure at maximum two varying policies per-each population. Theoretical justification can be found in ~\cite{fudenberg2006imitation}.}, $\boldsymbol{\pi} _{-i} = \boldsymbol{\pi} \setminus \pi _{i}$ representing policies followed by other agents, and $\alpha \in \mathbb{R}_{+}$ an ranking-intensity parameter. 
Large $\alpha$ ensures that the probability that a sub-optimal strategy overtakes a better strategy is close to zero.

As noted in~\cite{omidshafiei2019alpha}, one can relate the above switching process to a random walk on a Markov chain with states defined as elements in $\mathcal{S}_{\text{joint}} $. 
Essentially, the Markov chain models the \emph{sink strongly connected components} (SSCC) of the \emph{response graph} associated with the game. 
The response graph of a game is a directed graph where each node corresponds to each joint strategy profile, and directed edges if the deviating player's new strategy is a better response to that player, and the SSCC of a directed graph are the (group of) nodes with no out-going edges.

Each entry in the transition probability matrix   $\bm{T} \in \mathbb{R}^{\big|\mathcal{S}_{\text{joint}} \big| \times \big|\mathcal{S}_{\text{joint}} \big|}$ of Markov chain refers to the probability of one agent switching from one policy in a relation to attained payoffs. 
Consider any two joint strategy profiles $\pi_{\text{joint}} $ and $\hat{\pi}_{\text{joint}} $ that differ in only \emph{one} individual strategy for the $i^{th}$ agent, i.e., there exists an unique agent such that $\bm{\pi}_{\text{joint}}  = \big\{\pi_{i,a} , \bm{\pi_{-i}} \big\}$ and $\hat{\pi}_{\text{joint}} =\big\{\hat{\bm{\pi}}_{i,b} , \bm{\pi}_{-i} \big\}$ with $\pi_{i,a}  \neq \hat{\pi}_{i,b} $, we set 
$
    \left[\bm{T} \right]_{\bm{\pi}_{\text{joint}} , {\bm{\hat{\pi}}_{\text{joint}} }} = \frac{1}{\sum_{l=1}^{N} (k_{l} - 1)} \rho_{\pi_{i,a} , \hat{\pi}_{i,b} } \big(\bm{\pi_{-i}} \big),
$
with $\rho_{\pi_{i,a} , \hat{\pi}_{i,b} } \big(\bm{\pi_{-i}} \big)$ defining the probability that one copy of agent $i$ with strategy $\pi_{i,a} $ invades the population with all other agents (in that population) playing $\hat{\pi}_{i, b} $. Following \cite{pinsky2010introduction}, for $\mathcal{P}_{i} \big(\pi_{i,a} , \bm{\pi}_{-i} \big) \neq \mathcal{P}_{i} \big(\hat{\pi}_{i,b} , \bm{\pi}_{-i} \big)$, such a probability is formalised as 
\begin{align}
    \rho_{\pi _{i,a},\hat{\pi} _{i,b}}(\boldsymbol{\pi} _{-i}) = \frac{1 - e^{-\alpha\left(\mathcal{P} _i(\pi _{i,a},\boldsymbol{\pi} _{-i}) - \mathcal{P} _i(\hat{\pi} _{i,b},\boldsymbol{\pi} _{-i})  \right)}}{1 - e^{-m\alpha\left(\mathcal{P} _i(\pi _{i,a},\boldsymbol{\pi} _{-i}) - \mathcal{P} _i(\hat{\pi} _{i,b},\boldsymbol{\pi} _{-i})  \right)}},  \label{transition_prob}
\end{align}
and $\frac{1}{m}$ otherwise, with $m$ being the size of the population. So far, we presented relevant derivations for the $(\bm{\pi}_{\text{joint}} , \hat{\bm{\pi
}}_{\text{joint}} )$ entry of the state transition matrix when exactly the $i^{th}$ agent differs in exactly one strategy. Having one policy change, however, only represents a subset of allowed variations; two more cases need to be considered. Now we restrict our attention to variations in joint policies involving more than two individual strategies, i.e., $\big|\bm{\pi}_{\text{joint}}  \setminus \hat{\bm{\pi}}_{\text{joint}}  \geq 2 \big|$. Here, we set\footnote{
This assumption significantly reduces the analysis complexity as detailed in~\cite{fudenberg2006imitation}.} $\left[\bm{T} \right]_{\bm{\pi}_{\text{joint}} , \hat{\bm{\pi}}_{\text{joint}} } = 0$. 
Consequently, the remaining event of self-transitions can be thus written as  $\left[\bm{T} \right]_{\bm{\pi}_{\text{joint}} , \hat{\bm{\pi}}_{\text{joint}} } = 1 - \sum_{\hat{\bm{\pi}} }\left[\bm{T} \right]_{\bm{\pi}_{\text{joint}} , \hat{\bm{\pi}} }$. Summarising the above three cases, we can  write the $(\bm{\pi}_{\text{joint}} , \hat{\bm{\pi
}}_{\text{joint}} )$'s entry of the Markov chain's transition matrix as: 
{\normalsize
\begin{equation}
    \left[\bm{T} \right]_{\bm{\pi} _{\text{joint}}, \hat{\bm{\pi}} _{\text{joint}}} = \left\{\begin{aligned}
         & \dfrac{1}{\sum_{l=1}^{N} (k_{l} - 1)} \rho_{\pi_{i,a} , \hat{\pi}_{i,b} } \big(\bm{\pi_{-i}} \big), \ \ \text{if  $|\bm{\pi}_{\text{joint}}  \setminus \hat{\bm{\pi}}_{\text{joint}}  |=1$,} \\
         & 1 - \sum_{\hat{\bm{\pi}}\neq \bm{\pi}_{\text{joint}}}\left[\bm{T} \right]_{\bm{\pi}_{\text{joint}} , \hat{\bm{\pi}} }, \ \   \text{if $\bm{\pi}_{\text{joint}}  = \hat{\bm{\pi}}_{\text{joint}} $,} 
         \\ 
         & 0, \ \ \ \ \ \ \ \ \   \text{if $|\bm{\pi}_{\text{joint}}  \setminus \hat{\bm{\pi}}_{\text{joint}}   |\geq 2$,} \label{Eq:Transition} 
         \end{aligned}
    \right.
\end{equation}}
\begin{algorithm}[t!]
 \caption{$\alpha$-Rank (see Section  3.1.1 in \cite{omidshafiei2019alpha})}\label{algo:arank}
 \begin{algorithmic}[1]
 \State \textbf{(Unspecified) Inputs: } $\mathcal{S}_{\text{joint}}$, Multi-agent Simulator 
  \vspace{2pt}
  \State Listing all possible joint-strategy profiles,  for each profile, run the multi-agent simulator to get the payoff values for all players $\mathcal{P}_{\text{joint}} =\big\{\mathcal{P}_{1} \big(\pi_{\text{joint}} \big), \dots, \mathcal{P}_{N} \big(\pi_{\text{joint}} \big)\big\}, \forall \pi_{\text{joint}} \in \mathcal{S}_{\text{joint}}$.
  \vspace{2pt}
  \State Construct Markov chain's transition matrix $\bm{T}$ by  Eqn.~\ref{Eq:Transition}.
    \vspace{2pt}
   \State Compute the stationary distribution $\bm{v}$  by   Eqn.~\ref{Eq:Eign}.
     \vspace{2pt}
   \State Rank all $\pi_{\text{joint}}$ in $\bm{v}$ based on their probability masses.
        \vspace{2pt}
 \State \textbf{Outputs:} The ranked list of $\bm{v}$ (each element refers to the time that players spend in playing that $\pi_{\text{joint}}$ during evolution).  
\end{algorithmic}
 \end{algorithm}
The goal in $\alpha$-Rank is to establish an ordering in policy profiles dependent on evolutionary stability of each joint strategy. In other words, higher ranked strategies are these that are prevalent in populations with higher average time of survival. Formally, such a notion can be easily derived as the limiting vector $\bm{v}  = \lim_{t\rightarrow \infty}  \left[\left[\bm{T} \right]^{T}\right]^{t}\bm{v}_{0}$ 
of our Markov chain when evolving from an initial distribution $\bm{v}_{0}$. Knowing that the limiting vector is a stationary distribution, one can calculate in fact the solution to the following eigenvector problem: 
\begin{equation}
\label{Eq:Eign}
    \left[\bm{T} \right]^{T} \bm{v}  = \bm{v} . 
\end{equation}
We summarised the pseudo-code of $\alpha$-Rank  in Algorithm~\ref{algo:arank}. 
As the input to $\alpha$-Rank is unclear and  turns out to be controversial later, we point the readers to  the original  description in Section  3.1.1 of \cite{omidshafiei2019alpha}, and the practical implementation of $\alpha$-Rank from \cite{lanctot2019openspiel} for self-judgement.
In what comes next, we demonstrate that the tractability claim of $\alpha$-Rank needs to be relaxed as the algorithm exhibits exponential time and memory complexities in number of players dependent on the input type considered. 
This, consequently, renders $\alpha$-Rank inapplicable to large-scale multi-agent systems contrary to the original presentation.

\section{Claims \& Refinements}\label{Sec:Three}
Original presentation of $\alpha$-Rank claims to be tractable in the sense that it runs in polynomial time with respect to the total number of joint-strategy profiles. Unfortunately, such a claim is not clear without a formal specification of the inputs to Algorithm 3.1.1 in \cite{omidshafiei2019alpha}. In fact, we, next, demonstrate that $\alpha$-Rank's can easily exhibit exponential complexity under 
the input of $N\times k$ table, rendering it inapplicable beyond finite small number of players. We also present a conjecture stating that determining the top-rank joint strategy profile in $\alpha$-Rank is in fact NP-hard. 

\subsection{On $\alpha$-Rank's Computational Complexity}
Before diving into the  details of our arguments, it is first instructive to note that \emph{tractable} algorithms are these that exhibit a worst-case polynomial running time in the size of their input \cite{papadimitriou2003computational}. 
Mathematically, for a size $\mathcal{I}$ input, a polynomial time algorithm adheres to an $\mathcal{O}(\mathcal{I}^{d})$ complexity for some constant $d$ independent of $\mathcal{I}$. 

Following the presentation in Section 3.1.1 in~\cite{omidshafiei2019alpha}, $\alpha$-Rank assumes availability of a game simulator to construct a payoff matrix quantifying performance of joint strategy profiles. As such, we deem that necessary inputs for such a construction is of the size  $\mathcal{I} = N \times k$, where $N$ is the total number of agents and $k$ is the total number of strategies per agent, where we assumed $k=k_i=k_j$ for simplicity. 

Following the definition above, if $\alpha$-Rank possesses polynomial complexity then it should attain a time proportional to $\mathcal{O}\left(\left(N \times k\right)^d\right)$ with $d$ being a constant independent of $N$ and $k$. As the algorithm requires to compute a stationary distribution of a Markov chain described by a transition matrix $\bm{T}$ with $k^{N}$ rows and columns, the time complexity of $\alpha$-Rank amounts to $\mathcal{O}\left(k^{N}\right)$. Clearly, this result demonstrates exponential, thus intractable, complexity in the number of agent $N$. In fact, we conjecture that determining top rank joint strategy profile using $\alpha$-Rank with an $N\times k$ input is NP-hard.

\begin{conjecture}[$\alpha$-Rank is \textbf{NP-hard}]
\emph{Consider $N$ agents each with $k$ strategies. Computing top-rank joint strategy profile with respect to the stationary distribution of the Markov chain's transition matrix, $\bm{T}$, is \textbf{NP-hard}}.   
\end{conjecture}

\noindent\underline{\textbf{Reasoning:}} To illustrate the point of the conjecture above, imagine $N$ agents each with $k$ strategies. Following the certificate argument for determining complexity classes, we ask the  question: 
\begin{center}
    \emph{``Assume we are given a joint strategy profile $\pi_{\text{joint}}$, 
    is $\pi_{\text{joint}}$ top rank w.r.t the stationary distribution of the Markov chain?"}
\end{center}
To determine an answer to the above question, one requires an evaluation mechanism of some sort. If the time complexity of this mechanism is polynomial with respect to the input size, i.e., $N \times k$, then one can claim that the problem belongs to the NP complexity class. 
However, if the aforementioned mechanism exhibits an exponential time complexity, then the problem belongs to NP-hard complexity class. When it comes to $\alpha$-Rank, we believe a mechanism answering the above question would require computing a holistic solution of the problem, which, unfortunately, is exponential (i.e., $\mathcal{O}(k^{N})$). Crucially, if our conjecture proves correct, we do not see  how $\alpha$-Rank can handle more than a finite small number of agents. 
\qed

\subsection{On Optimisation-Based Techniques}
Given exponential complexity as derived above, we can resort to approximations of stationary distributions that aim at determining $\epsilon$-close solution for some precision parameter $\epsilon > 0$. Here, we note that a problem of this type is a long-standing classical problem from linear algebra. Various techniques including Power method, PageRank, eigenvalue decomposition, and mirror descent can be utilised. Briefly surveying this literature, we demonstrate that any such implementation (unfortunately) scales exponentially in the number of players. For a quick summary, please consult Table~\ref{table:complexity}.

\subsubsection*{\textbf{Power Method.}} One of the most common approaches to computing a stationary distribution is the power method that computes the stationary vector $\bm{v}$ by constructing a sequence $\left\{\bm{v}_{j}\right\}_{j \geq 0}$ from a non-zero initialisation $\bm{v}_{0}$ by applying $\bm{v}_{j+1} = \sfrac{1}{\left|\left|\bm{T}^{, \mathsf{T}}\bm{v}_{j}\right|\right|}\bm{T}^{ \mathsf{T}}\bm{v}_{j}$. Though viable, we first note that the power method exhibits an exponential memory complexity in terms of the number of agents. 
To formally derive the bound, define $n$ to represent the total number of joint strategy profiles, i.e., $n=k^N$, and $m$ the total number of transitions between the states of the Markov chain. By construction, one can easily see that $m  = n \big( k N - N + 1\big)$ as each row and column in $\bm{T}$ contains $k N - N + 1$ non-zero elements. Hence, memory  complexity of such  implementation is in the order of $$\mathcal{O}(m) = \mathcal{O}\big(n\big[ k N - N + 1\big]\big) \approx \mathcal{O}\big( k^{N} k N \big).$$

The time complexity of a power method, furthermore, is given by $\mathcal{O}(m \times \mathcal{T})$, where $\mathcal{T}$ is the total number of iterations. Since $m$ is of the order $n\log n$, the total complexity of such an implementation is also exponential.

\subsubsection*{\textbf{PageRank.}} Inspired by ranking web-pages on the internet, one can consider PageRank~\cite{page1999pagerank} for computing the solution to the eigenvalue problem presented above. Applied to our setting, we first realise that the memory is analogous to the power method that is $\mathcal{O}(m)  = \mathcal{O}\big(K^{N+1} N \big)$, 
 and the time complexity are in the order of $\mathcal{O}(m + n) \approx \mathcal{O}\big(K^{N+1} N\big)$.
 
\begin{table}[t!]
\large{
\caption{Time and space complexity comparison given $N\text{(number of agents)} \times k\text{(number of strategies)}$ table as inputs.}
\label{table:complexity}
\vspace{-10pt}
\begin{center}
  \begin{tabular}{| c || c | c | }
    \hline 
    Method & $\text{ Time }$ & $\text{ Memory }$ \\  \hline 
    $\text{ Power Method }$ & $\mathcal{O}\left(k^{N+1}N\right)$ & $\mathcal{O}\left(k^{N+1}N \right)$ \\ \hline
    $\text{ PageRank }$ & $\mathcal{O}\left(k^{N+1}N\right)$ & $\mathcal{O}\left(k^{N+1}N \right)$ \\ \hline
    $\text{ Eig. Decomp. }$ & $\mathcal{O}\left(k^{N\omega}\right)$ & $\mathcal{O}\left(k^{N+1} N\right)$ \\ \hline
    $\text{ Mirror Descent }$ & $\mathcal{O}\left(k^{N+1}\log k \right)$ & $\mathcal{O}\left(k^{N+1} N\right)$ \\ \hline
  \end{tabular}
\end{center}
}
\vspace{-0pt}
\end{table}

\subsubsection*{\textbf{Eigenvalue Decomposition.}}
Apart from the above, we can also consider the problem as a standard eigenvalue decomposition task (also what is used to implement $\alpha$-Rank in \cite{lanctot2019openspiel}) and adopt the method in~\cite{coppersmith1990matrix} to compute the stationary distribution. Unfortunately, state-of-the-art techniques for eigenvalue decomposition also require exponential memory ($\mathcal{O}\left(K^{N+1} N\right)$) and exhibit a time complexity of the form $\mathcal{O}(n^{\omega}) = \mathcal{O}(k^{N\omega})$ with $\omega \in [2, 2.376]$ \cite{coppersmith1990matrix}. Clearly, these bounds restrict $\alpha$-Rank to small agent number $N$.

\subsubsection*{\textbf{Mirror Descent.}}
 Another optimisation-based alternative is the ordered subsets mirror descent algorithm \cite{ben2001ordered}. This is an iterative procedure requiring projection step on the standard $n$-dimensional simplex on every iteration:
$
    \bm{\Delta}_n = \{\boldsymbol{x}\in\mathbb{R}^n: \boldsymbol{x}^{\mathsf{T}}\boldsymbol{1} = 1 \ \& \ \ \boldsymbol{x} \succeq \boldsymbol{0}\}.
$
As mentioned in~\cite{ben2001ordered}, computing this projection requires $\mathcal{O}(n\log n)$ time. Hence, the projection step is exponential in the number of agents $N$. This makes mirror descent  inapplicable to $\alpha$-Rank when $N$ is large. 

Apart from the methods listed above, we are aware of other approaches that could  solve the leading eigenvector for big matrices, for example the online learning  approach \cite{garber2015online}, the sketching methods \cite{tropp2017practical}, and the subspace iteration with Rayleigh-Ritz acceleration \cite{golub2000eigenvalue}. The trade-off of these methods is that they usually assume special structure of the matrix, such as being Hermitian  or at least positive semi-definite, which $\alpha$-Rank however does not fit. Importantly, they can not offer any advantages on the time  complexity either.

\section{Reconsidering $\alpha$-Rank's Inputs}
Having discussed our results with the authors, we were suggested that ``inputs'' to $\alpha$-Rank are \textbf{exponentially-sized} payoff matrices, i.e., assuming line 2 in Algorithm~\ref{algo:arank} as an input 
. Though polynomial in an exponentially-sized input, this consideration does not resolve problems mentioned above. In this section, we further demonstrate additional theoretical and practical problems when considering the advised ``input" by the authors.

\subsection{On the Definition of \emph{Agents}}\label{Sec:Alpha} $\alpha$-Rank redefines a strategy to correspond to the agents under evaluation differentiating them from players in the game (see line 4 in Section 3.1.1 and also  Fig. 2a in \cite{omidshafiei2019alpha}). 
Complexity results are then given in terms of these ``agents", where tractability is claimed. We would like to clarify that such definitions do not necessarily reflect the true underlying time complexity, whereby without formal input definitions, it is difficult to claim tractability. 

To illustrate, consider solving a travelling salesman problem in which a traveller needs to visit a set of cities while returning to the origin following the shortest route. Although it is well-known that the travelling salesman problem is NP-hard, following the line of thought presented in $\alpha$-Rank, one can show that such a problem reduces to a polynomial time (linear, i.e., tractable) problem in the size of ``meta-cities'', which is not a valid claim.
\begin{center}
\textbf{So what are the ``meta-cities'', and what is \\wrong with the above argument?}    
\end{center}
A strategy in the travelling salesman problem corresponds to a permutation in the order of cities. Rather than operating with number of cities, following $\alpha$-Rank, we can construct the space of all permutation calling each a ``meta-city'' (or agent)\footnote{How to enumerate all these permutations of cities is analogous to  enumerating an exponentially sized matrix in $\alpha$-Rank if $N \times k$ was not the input to $\alpha$-Rank.}. Having enumerated all permutations, somehow, searching for the shortest route can be performed in polynomial time. Even though, one can state that solving the travelling salesman problem is polynomial in the size of permutations, it is incorrect to claim that any such algorithm is tractable. The same exact argument can be made for $\alpha$-Rank, whereby having a polynomial time algorithm in an exponentially-sized space does not at all imply tractability\footnote{Note that this claim does not apply on the complexity of solving Nash equilibrium. For example, in solving zero-sum games, polynomial tractability is never claimed on the number of players, whereas $\alpha$-Rank claims tractable in the number of players.}. It is for this reason, that reporting complexity results needs to be done with respect to the size of the input without any redefinition (we believe these are agents in multi-agent systems, and cities in the travelling salesman problem).   
\begin{figure}[t!]
    \centering
        \vspace{-25pt}
    \subfigure{\label{fig:dollar}\includegraphics[trim = {10em, 8em, 5em, 5em}, width=.88\columnwidth]{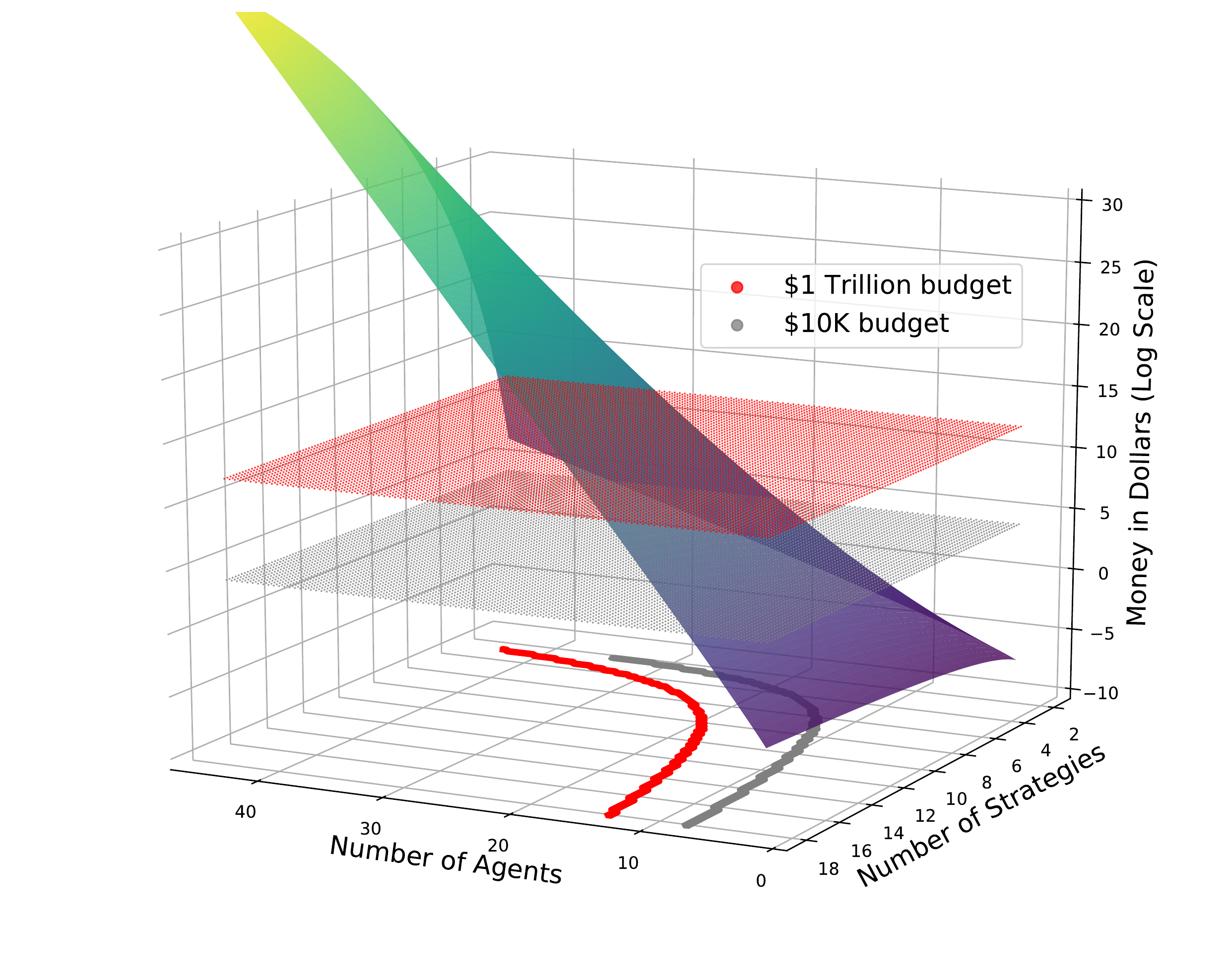}}
    \vspace{-5pt}
    \caption{Money cost of constructing the transition matrix $\bm{T}$ in computing $\alpha$-Rank (line 3 in Algorithm~\ref{algo:arank}). Note that one trillion dollar is the world's total hardware budget \cite{moneybudget}.  The projected contours shows that due to the exponentially-growing size of $\alpha$-Rank's "input",  under reasonable budget, it is infeasible to handle multi-agent evaluations with more than ten agents. 
    }\label{fig:money}
        \vspace{-0pt}
\end{figure}

As is clear so-far, inputs to $\alpha$-Rank lack clarity. Confused on the form of the input, we realise that the we are left with two choices: 1) list of all joint strategy profiles, or 2) a table of the size $N \times k$ -- collection of all of the players' strategy pools. If we are to follow the first direction, the claims made in the paper are of course correct; however, this by no means resolves the problem as it is not clear how one would construct such an input in a tractable manner. Precisely, given an $N \times k$ table (collection of all of the players' strategy pools) as input, constructing the aforementioned list requires exponential time ($k^{N}$). In other words, providing $\alpha$-Rank with such a list only hides the exponential complexity burden in a pre-processing step. Analogously, applying this idea to the travelling salesman problem described above would hide the exponential complexity under a pre-processing step used to construct all possible permutations. Provided as inputs, the travelling salesman problem can now be solved in linear time, i.e., transforming an intractable problem to a tractable one by a mere redefinition.

\begin{table}[t!]
\begin{center}
\vspace{-10pt}
\caption{Cost 
of getting the payoff table $\mathcal{P}_{\text{joint}}$ (line 2 in Algorithm~\ref{algo:arank})  for the experiments conducted in \citet{omidshafiei2019alpha}. We list the numbers by the cost of running one joint-strategy profile $\times$ the number of joint-strategy profiles considered. 
Detailed computation can be found \href{https://drive.google.com/open?id=1XJM4C9WWRJTBrIzSuBkpAnleoo3B9TanMCl7nWVIn_s}{here.}}
\vspace{-10pt}
\resizebox{1.\columnwidth}{!}{
\begin{tabular}{|l||c|c|c|}
\hline
{\textbf{Game Env.}} & {\textbf{PetaFlop/s-days}} & \textbf{Cost (\$)}   & \textbf{Time (days)}   \\ \hline
{AlphaZero Go~\citep{silver2017mastering}}              & {$1,413 \times 7$}                      & {$207M$}       &      {$1.9M$}                                       \\ \hline
{AlphaGo Zero
~\citep{silver2016mastering}}           & {$1,181 \times 7$}                      & {$172M$}                    &     {$1.6M$}                           \\ \hline
{AlphaZero Chess~\citep{silver2017mastering}}              & {$17 \times 1$}                      & {$352K$}       &      {$3.2K$}                                       \\ \hline
{MuJoCo Soccer~\citep{liu2019emergent}}   & {$0.053 \times 10$}                      & $4.1K$                       &       $72$                     \\ \hline
{Leduc Poker~\citep{lanctot2017unified}}   & {$0.006 \times 9$}                      & {$420$}                    &     $7$                         \\ \hline 
{Kuhn Poker~\citep{heinrich2015fictitious}}   & {$<10^{-4} \times 256$}                      & {$<1$}                    &     $-$                         \\ \hline \hline
{AlphaStar~\citep{vinyals2019grandmaster}}              & $52,425$                      & {$244M$}       &      {$1.3M$}                                       \\ \hline 
\end{tabular} \label{table:money}
}
\end{center}
\end{table}
\subsection{Dollars Spent: A Non-Refutable Metric}
Admittedly, our arguments have been mostly theoretical and can become controversial dependent on the setting one considers. To abolish any doubts, we followed the advice given by the authors and considered the input of $\alpha$-Rank to be exponentially-sized payoff matrices. We then conducted an experiment measuring dollars spent to evaluate scalability of running just line 3 in Algorithm~\ref{algo:arank}, while considering the tasks reported in~\cite{omidshafiei2019alpha}.   

Assuming $\mathcal{P}_{\text{joint}} =\big\{\mathcal{P}_{1} \big(\pi_{\text{joint}} \big), \dots, \mathcal{P}_{N} \big(\pi_{\text{joint}} \big)\big\}, \forall \pi_{\text{joint}} \in \mathcal{S}_{\text{joint}}$ is given at no cost, the total amount of floating point operations (FLOPS) needed for constructing $\bm{T}$ given in Eqn.~\ref{Eq:Transition} is $9k^NN(k-1) + k^{N}N(k-1) + 0 = 10k^{N}N(k-1)$.
In terms of money cost needed for just building $\bm{T}$, we plot the dollar amount in Fig.~\ref{fig:money} considering the Nvidia Tesla K80 GPU\footnote{\url{https://en.wikipedia.org/wiki/Nvidia_Tesla}} which can process under single precision at maximum $5.6$ TFlop/s at a price of $0.9 \  \$$/hour\footnote{\url{https://aws.amazon.com/ec2/instance-types/p2/}}.
 Clearly, Fig.~\ref{fig:money} shows that due to the fact that $\alpha$-Rank needs to construct a Markov chain with an exponential size in the number of agents, it is only ``money'' feasible on tasks with at most tens of agents. It is also worth noting that our analysis is optimistic in the sense that we have not considered costs of storing $\bm{T}$ nor computing stationary distributions.  
 
 \vspace{2pt}
 \noindent\underline{\textbf{Conclusion I:}} \emph{Given exponentially-sized payoff matrices, constructing transition matrices in $\alpha$-Rank for about $20$ agents each with $8$ strategies requires about \textbf{one trillion dollars} in budget.}
 \vspace{2pt}

Though assumed given, in reality, the payoff values $\mathcal{P}_{\text{joint}}$ come at a non-trivial cost themselves, which is particularly true in reinforcement learning tasks \cite{silver2016mastering}.
Here, we take a closer look at the amount of money it takes to attain payoff matrices for the experiments listed in \cite{omidshafiei2019alpha} that we present in Table \ref{table:money}. 
Following the methodology in \href{https://openai.com/blog/ai-and-compute/}{here}, we first count the total FLOPS each model uses under the unit of PetaFlop/s-day that consists of performing $10^{20}$ operations per second in one day. 
For each experiment, if the answer to ``how many GPUs were trained and for how long'' was not available, we then traced back to the neural architecture used and counted the operations needed for both forward and backward propagation. 
The cost in time was then transformed from PetaFlop/s-day using Tesla K80 as discussed above.
In addition, we also list the cost of attaining payoff values from the most recent AlphaStar model \cite{vinyals2019grandmaster}.
It is obvious that although $\alpha$-Rank could take the payoff values as ``input'' at a hefty price, the cost of acquiring such values is not negligible, e.g., payoff values from GO cost about $207M$ \$, and require a single GPU to run for more than five thousand years\footnote{
Note that here we only count running experiment once for getting each payoff value. In practice,   the game outcomes are noisy, multiple samples are often  needed  (check Theorem 3.2 in \cite{rowl2019multiagent}), which will turn the numbers in Table \ref{table:money} to an even larger scale. }! 

 \vspace{2pt}

\noindent\underline{\textbf{Conclusion II:}} \emph{Acquiring necessary inputs to $\alpha$-Rank easily becomes intractable giving credence to our arguments in Section~\ref{Sec:Alpha}.}

\section{A Practical Solution to $\alpha$-Rank}
\label{Sec:Scale}
One can consider approximate solutions to the problem in Eqn.~\ref{Eq:Eign}. As briefly surveyed in Section \ref{Sec:Three}, most current methods, unfortunately, require exponential \emph{time and memory} complexities. We believe achieving a solution that aims at reducing time complexity is an interesting and open question in linear algebra in general, and leave such a study to future work. Here, we rather contribute by a stochastic optimisation method that can attain a solution through random sampling of payoff matrices without the need to store exponential-size input. Contrary to memory requirements reported in Table \ref{table:complexity}, our method requires a linear (in number of agents) per-iteration complexity of the form $\mathcal{O}(Nk)$. It is worth noting that most other techniques need to store exponentially-sized matrices before commencing with any numerical instructions. Though we do not theoretically contribute to reductions in time complexities, we do, however, augment our algorithm with a double-oracle heuristic for joint strategy space reduction. In fact, our experiments reveal that $\alpha^{\alpha}$-Rank can converge to the correct top-rank strategies in hundreds of iterations in large strategy spaces, i.e., spaces with $\approx$ 33 million profiles.        

\textbf{Optimisation Problem Formulation:} Computing the  stationary distribution can be rewritten as an optimisation problem: 
\begin{equation}
\label{Eq:OptimObj}
\min_{\bm{x}} \frac{1}{n} \left|\left|\bm{T}^{   T}\bm{x} - \bm{x}\right|\right|_{2}^{2} \ \ \text{s.t. $\bm{x}^{T}\bm{1} = 1$, and $\bm{x} \succeq \boldsymbol{0}$,}
\end{equation}
where the constrained objective in Eqn.~\ref{Eq:OptimObj} simply seeks a vector $\bm{x}$ minimising the distance between $\bm{x}$, itself, and $\bm{T}^{   T}\bm{x}$  while ensuring that $\bm{x}$ lies on an $n$-dimensional. To handle exponential complexities needed for acquiring exact solutions, we pose a relaxation the problem in Eqn.~\ref{Eq:OptimObj} and  focus on computing an approximate solution vector $\tilde{\bm{x}}$ instead, where $\tilde{\bm{x}}$ solves: 
{\normalsize 
\begin{equation}
\label{Eq:OptimObjRelaxed}
\min_{\bm{x}} \frac{1}{n} \left|\left|\bm{T}^{   T}\bm{x} - \bm{x}\right|\right|_{2}^{2} \ \ \text{s.t. $\left|\bm{x}^{T}\bm{1} - 1\right| \leq \delta$ for $0 < \delta < 1$, and $\bm{x} \succeq \boldsymbol{0}$.}
\end{equation}}
Before proceeding, however, it is worth investigating the relation between the solutions of the original (Eqn. \ref{Eq:OptimObj}) and relaxed (Eqn. \ref{Eq:OptimObjRelaxed}) problems. We summarise such a relation in the following proposition that shows that determining $\tilde{\bm{x}}$ suffices for computing a stationary distribution of $\alpha$-Rank's Markov chain:

\begin{proposition}[Connections to Markov Chain]
\emph{Let $\tilde{\bm{x}}$ be a solution to the relaxed optimisation problem in Eqn.~\ref{Eq:OptimObjRelaxed}. Then, $\sfrac{\tilde{\bm{x}}}{||\tilde{\bm{x}}||_{1}} = \bm{v}^{ }$ is the stationary distribution of Eqn.~\ref{Eq:Eign} in Section~\ref{Sec:MC}}.
\end{proposition}

Importantly, the above proposition, additionally, allows us to focus on solving the problem in Eqn.~\ref{Eq:OptimObjRelaxed} that only exhibits inequality constraints. Problems of this nature can be solved by considering a barrier function leading to an unconstrained finite sum minimisation problem. By denoting $\bm{b}_{i}^{ }$ to be the $i^{th}$ row of $\bm{T}^{   T} - \bm{I}$, we can, thus, write:  
$
    \frac{1}{n}\left|\left|\bm{T}^{   T} \bm{x} - \bm{x} \right|\right|_{2}^{2} = \frac{1}{n} \sum_{i=1}^{n} \left(\bm{x}^{T}\bm{b}_{i}^{ }\right)^{2}. 
$ Introducing logarithmic barrier-functions, with $\lambda > 0$ being a penalty parameter, we have:    
{\small
\begin{equation}
\label{Eq:LogBarrier}
    \min_{\boldsymbol{x}\in\mathbb{R}^n}  \frac{1}{n}\sum_{i=1}^n\left(\boldsymbol{x}^{T}\boldsymbol{b}^{ }_i\right)^2 - \lambda\log\left(\delta^2 - \left[\boldsymbol{x}^{T}\boldsymbol{1} - 1\right]^2\right) - \frac{\lambda}{n}\sum_{i=1}^{n}\log(x_i).
\end{equation}
}
 Eqn.~\ref{Eq:LogBarrier} represents a standard finite minimisation problem, which can be solved using any off-the-shelf stochastic optimisation methods, e.g., stochastic gradients, ADAM \cite{kingma2014adam}. A stochastic gradient execution involves sampling a strategy profile $i_{t}\sim[1, \dots, n]$ at iteration $t$, and then executing a descent step: $ \bm{x}_{t+1} = \bm{x}_{t} - \eta_{t} \nabla_{\bm{x}} f_{i_{t}}(\bm{x}_{t})$,
with $\nabla_{\bm{x}} f_{i_{t}}(\bm{x}_{t})$ being a sub-sampled gradient of  Eqn.~\ref{Eq:LogBarrier}, and $\lambda$ being a scheduled penalty parameter with $\lambda_{t+1} = \sfrac{\lambda_{t}}{\gamma}$ for some $\gamma > 1$:
{\small\begin{align}
\label{Eq:Gradients}
    \nabla_{\bm{x}} f_{i_{t}}(\bm{x}_{t}) = 2 \left(\bm{b}_{i_{t}}^{   \bm{T}}\bm{1}\right)\bm{b}_{i_{t}}^{ } & +  \frac{2\lambda_{t}\left(\bm{x}_{t}^{T}\bm{1} - 1\right)}{\delta^{2} - \left(\bm{x}_{t}^{T}\bm{1} - 1\right)^{2}}  - \sfrac{\lambda_{t}}{n}\left[\frac{1}{[\bm{x}_{t}]_{1}}, \dots, \frac{1}{[\bm{x}_{t}]_{n}}\right]^{T}.
\end{align}}
\begin{algorithm}[t!]
 \caption{$\alpha^{\alpha}$-Oracle: Practical Multi-Agent Evaluation }\label{alpharank_alg}
 \begin{algorithmic}[1]
 \State \textbf{Inputs:} Number of trails $\mathcal{N}$, total number of iterations $T$, decaying learning rate $\left\{\eta_{t}\right\}_{t=1}^{T}$, penalty parameter $\lambda $, $\lambda$ decay rate $\gamma > 1$, and a constraint relaxation term $\delta$, initialise $p=0$.
 \State \textbf{while} $p \le \mathcal{N}$ \textbf{do}:
  \State \hspace{1em} Set the counter of running oracles, $k = 0$
  \State \hspace{1em} Initialise the strategy set $\{\mathcal{S}^{[0]}_i\}$ by sub-sampling from $\{\mathcal{S}_i \}$
   \State \hspace{1em} \textbf{while} $\{\mathcal{S}_{i}^{[k]}\} \ne \{\mathcal{S}_{i}^{[k-1]}\}$ \textbf{AND} $k \ge 1$  \textbf{do}:
     \State \hspace{2em} Compute total number of joint profiles $n=\prod_{i=1}^{N}|\mathcal{S}_i^{[k]}|$
  \State \hspace{2em} Initial a vector $\bm{x}_{0} = \sfrac{1}{n}\bm{1}$ 
 \State \hspace{2em} \textbf{for} $t = 0 \rightarrow T-1$ \textbf{do:}  \ \ \ \ \ \ \ \ \ $//$ $\bm{\alpha^\alpha}$\textbf{-Rank update} 
 \State \hspace{3em} Uniformly sample one strategy profile $i^{[k]}_{t} \sim \{1, \dots, n\}$ 
 \State \hspace{3em} Construct $b^{[k]}_{i_{t}}$ as the $i_{t}^{[k]}$ row of $\bm{T}^{[k], T} - \bm{I}$ 
 \State \hspace{3em} Update  $\bm{x}_{t+1}^{[k]} = \bm{x}_{t}^{[k]} - \eta_{t} \nabla_{\bm{x}}
 f_{i^{[k]}_{t}}(\bm{x}_{t}^{[k]})$ by Eqn.~\ref{Eq:Gradients} 
 \State \hspace{3em}  Set $\lambda_{t+1} = \lambda_{t}/\gamma$
 \State \hspace{2em} Get  $\bm{\pi}_{\text{joint}}^{[p], \text{top}}$ by ranking the prob. mass of $\bm{v}^{[k]}= \frac{\bm{x}^{[k]}_{T}}{||\bm{x}^{[k]}_{T}||_{1}}$ 
\State \hspace{2em} Set $k = k + 1$
 \State \hspace{2em} \textbf{for} each agent $i$ \textbf{do}:    \ \ \ \ \ \ \ \ \ \ $//$ \textbf{The oracles (Section \ref{Sec:ScaleLearn})}
 \State \hspace{3em} Compute the best response $\pi_i^*$  to  $\bm{\pi}_{\text{joint}}^{[p], \text{top}}$ by Eqn.~\ref{Eq:probH}
\State \hspace{3em} Update the strategy set by $\mathcal{S}_{i}^{[k]} = \mathcal{S}_{i}^{[k-1]}
\cup \pi_{i}^{\star}$
\State \hspace{1em}  Set $p = p + 1$
\State \textbf{Return:} The best performing joint-strategy profile $\bm{\pi}_{\text{joint}, \star}$ among $\{\bm{\pi}_{\text{joint}}^{[1:\mathcal{N}], \text{top}}\}$. 
\end{algorithmic}
 \end{algorithm}
To avoid any confusion, we name the above stochastic approach of solving $\alpha$-Rank via Eqn.~\ref{Eq:LogBarrier} \&~\ref{Eq:Gradients} as $\bm{\alpha^\alpha}$\textbf{-Rank} and present its pseudo-code in Algorithm \ref{alpharank_alg}. When comparing   
 our algorithm to these reported in Table \ref{table:complexity}, it is worth highlighting that computing updates using Eqn. \ref{Eq:Gradients} requires no storage of the full transition or payoff matrices as updates are performed only using sub-sampled columns as shown in line 11 in Algorithm \ref{alpharank_alg}.

  \begin{figure*}[t!]
    \centering
    \vspace{-25pt}
    \subfigure{    \includegraphics[width=.95\textwidth]{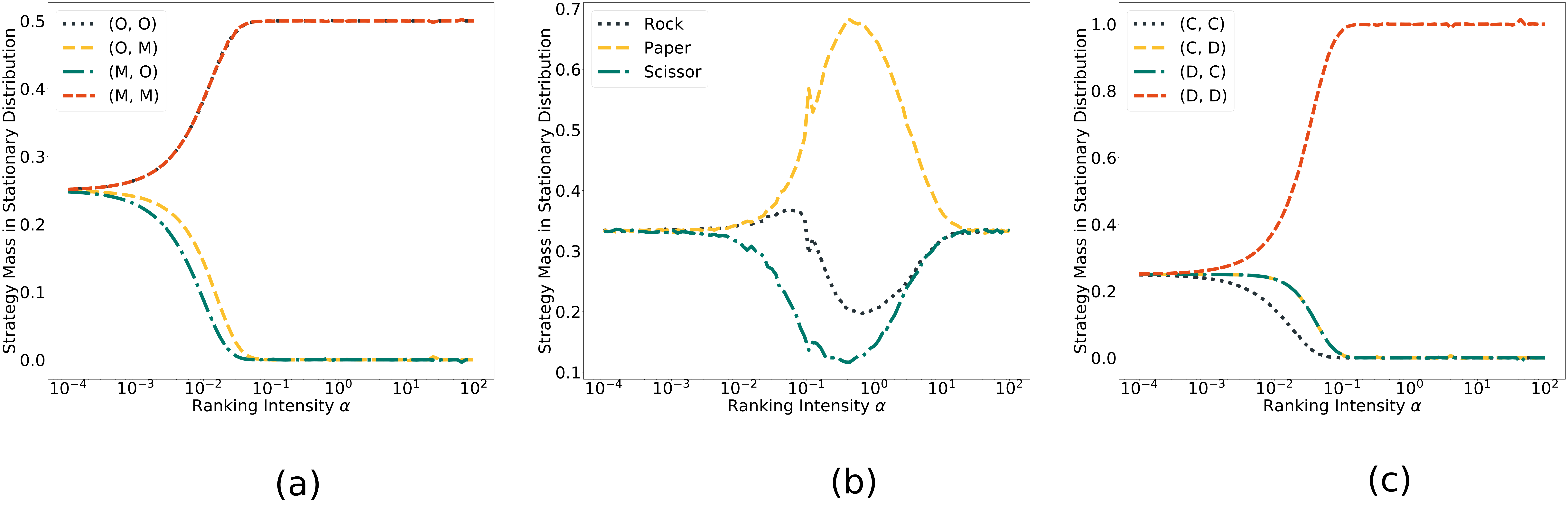}}
    \vspace{-17pt}
    \caption{Ranking intensity sweep on (a) Battle of Sexes  (b) Biased RPS  (c) Prisoner's Dilemma. 
    }
    \label{fig:3nfg}
    \vspace{-0pt}
\end{figure*}
 \begin{figure*}[t!]
    \centering
    \vspace{-10pt}
    \subfigure{\includegraphics[width=.8\textwidth]{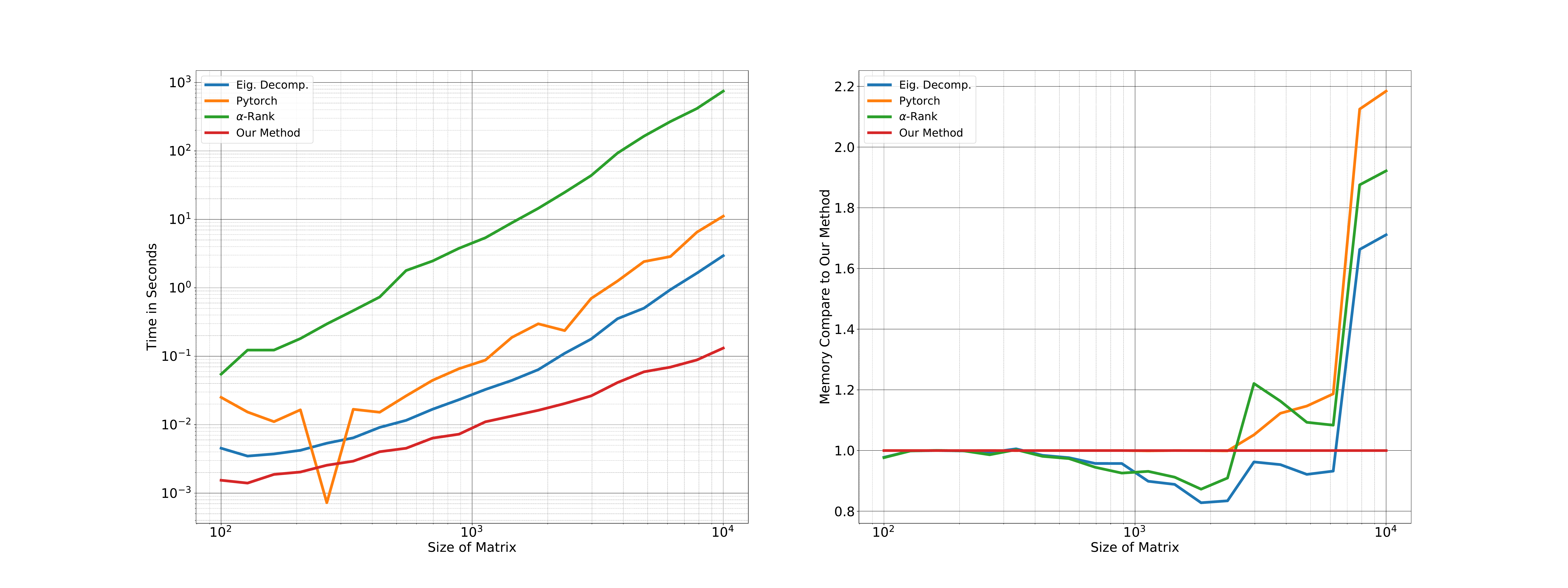}} 
    \vspace{-15pt}
    \caption{Comparisons of time and memory complexities on varying sizes of random matrices.
    }
    \label{fig:timemem}
    \vspace{-10pt}
\end{figure*}

\subsection{$\alpha^{\alpha}$-Rank with Efficient Exploration \& Oracles}\label{Sec:ScaleLearn}
Stochastic sampling enables to solve $\alpha$-Rank with no need to store the transition matrix $\bm{T}$; 
however, the size of the column $\bm{b}_{i}$ (i.e., $\prod_{i=1}^{N}k_i$) can still be prohibitively large. Here 
we further boost scalability of our method by introducing an \emph{oracle} mechanism.
The heuristic of oracles was first proposed in solving large-scale zero-sum matrix games \cite{mcmahan2003planning}. 
The idea is to first create a sub-game in which all players are only allowed to play a restricted number of strategies, which are then expanded by adding each of the players' best-responses to their opponents; the sub-game will be replayed with agents' augmented strategy sets before a new round of best responses is computed.

The best response is assumed to be given by an \emph{oracle} that can be simply implemented by a grid search, where given the top-rank profile $\pi_{-i}^{  \text{top}}$ at iteration $k$, the goal for agent $i$ is to  select the optimal $\pi_i^{*}$ from a pre-defined strategy set $\mathcal{S}_i$ to maximise its reward:  
\begin{equation}
\label{Eq:probH}
    \pi_{i}^{\star} = \arg\max_{\pi_{i}\in \mathcal{S}_i} \mathbb{E}_{\pi_{i},\pi_{-i}^{  \text{top}}}\Big[\sum_{h \geq 0} \gamma^{   h}_{i} \mathcal{P}_{i}^{ }(\bm{x}_{ h}^{ }, \bm{u}_{i, h}^{ }, \bm{u}_{-i, h}^{ })\Big], 
\end{equation}
with $\bm{x}_{ h}^{ }$ denoting the state, 
$\bm{u}_{i, h}^{ } \sim \pi_{i}(\cdot|\bm{x}_{i, h}^{ })$, $\bm{u}_{-i, h}^{ } \sim \pi_{-i}^{ \text{top}}(\cdot|\bm{x}_{-i, h}^{ })$ denoting the actions from agent $i$ and the opponents, respectively. Though worse-case scenario of introducing oracles would require solving the original evaluation problem, our experimental results on large-scale systems demonstrate efficiency by converging early. 

\begin{figure*}[t!]
    \centering
        \vspace{-25pt}
    \subfigure{\label{fig:a}\includegraphics[trim = {0em, 0em, 0em, 0em}, clip, width=1.\textwidth]{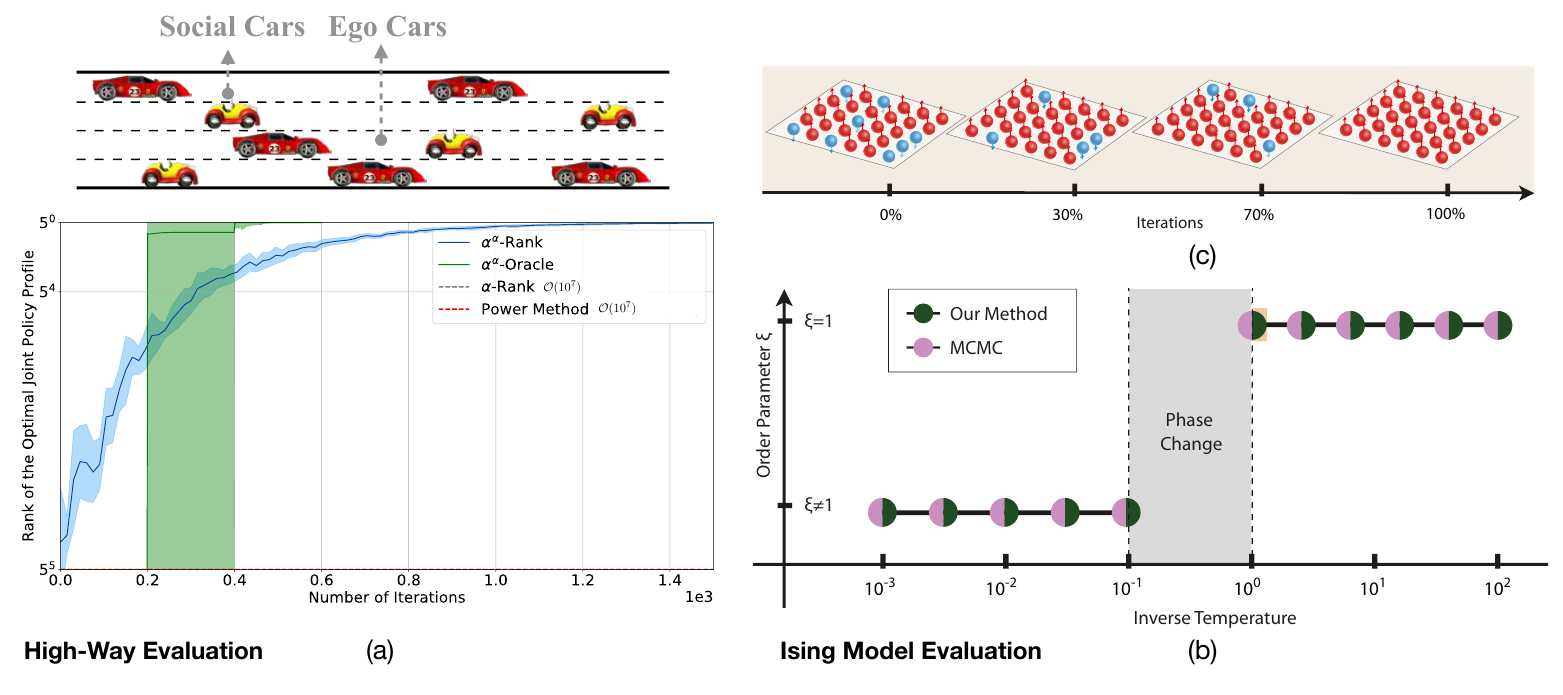}}
    \vspace{-22pt}
    \caption{Large-scale multi-agent evaluations.   (a)  Convergence of the optimal joint-strategy profile in  self-driving simulation. (b) Status of the Ising-model equilibrium measured by $\xi = \frac{|N_{\uparrow} - N_{\downarrow}|}{|N|}$. (c)  Change of the top-rank profile from $\alpha^\alpha$-Oracle under $\tau=1, \alpha=1$. 
    }
    \label{fig:ising_car}
        \vspace{-5pt}
\end{figure*}

For a complete exposition, we summarise the pseudo-code of our proposed method, named as $\bm{\alpha^\alpha}$\textbf{-Oracle}, in Algorithm 1.
$\alpha^\alpha$-Oracle degenerates to $\alpha^\alpha$-Rank (lines $6-13$) if one initialises strategy sets of agents by the full size at the beginning, i.e., $\{\mathcal{S}_i^{[0]}\} \triangleq \{\mathcal{S}_i\}$.

Providing valid convergence guarantee for $\alpha^\alpha$-Oracle is an interesting direction for future work. In fact, recently \cite{muller2019generalized} proposed a close idea of adopting an oracle mechanism into $\alpha$-Rank without any stochastic solver however. Interestingly, it is reported that bad initialisation can lead to failures in recovering top-rank strategies. Contrary to the results reported in  \cite{muller2019generalized}, we rather demonstrate the effectiveness of our approach through running multiple trails of initialisation for $\{\mathcal{S}_i^{[0]}\}$.
In addition, we also believe the stochastic nature of $\alpha^\alpha$-Oracle potentially prevents from being trapped by the local minimal from sub-games. 


\section{Experiments}\label{Sec:Exps}

In this section, we demonstrate the scalability of $\alpha^{\alpha}$-Rank in successfully recovering optimal policies in self-driving car simulations and in the Ising model--a setting with tens-of-millions of possible strategies. We note that these sizes are far beyond the capability of state-of-the-art methods; 
 $\alpha$-Rank \cite{omidshafiei2019alpha} considers at maximum $4$ agents with $4$ strategies. 
  All of our experiments were run only on a single machine with $64$ GB memory and $10$-core Intel i9 CPU.

\textbf{{Sparsity Data Structure:}} During the implementation phase, we realised that the transition probability, $\bm{T}^{[k]}$, of the Markov chain induces a sparsity pattern (each row and column in $\bm{T}^{[k]}$ contains $\sum_{i=1}^{N}k_{i}^{[k]} - N + 1$ non-zero elements, check Section~\ref{Sec:Scale}) that if exploited can lead to significant speed-up. To fully leverage such sparsity, we tailored a novel data structure for sparse storage and computations needed by Algorithm~\ref{alpharank_alg}. More details
are in \href{https://drive.google.com/file/d/1xz3OMSM8uUc3DpHCke-gqXq_Z37cb_ea/view}{Appendix}.

\textbf{{Correctness of Ranking Results:}}
 As Algorithm~\ref{alpharank_alg} is a generalisation (in terms of scalability) of $\alpha$-Rank, 
 it is instructive to validate the correctness of our results on three simple matrix games. Due to space constraints, we refrain the full description of these tasks to \href{https://drive.google.com/file/d/1xz3OMSM8uUc3DpHCke-gqXq_Z37cb_ea/view}{Appendix}. Fig.~\ref{fig:3nfg}, however,  shows that, in fact, results generated by $\alpha^\alpha$-Rank are consistent with these reported in \cite{omidshafiei2019alpha}. 

\textbf{{{Complexity Comparisons on Random Matrices:}}} To further assess scalability, we measured the time and memory needed by our method for computing stationary distributions of varying sizes of simulated random matrices. Baselines included eigenvalue decomposition from Numpy, optimisation tools from PyTorch, and $\alpha$-Rank from OpenSpiel \cite{lanctot2019openspiel}. We terminated execution of $\alpha^{\alpha}$-Rank when gradient norms fell-short a predefined threshold of \textbf{0.01}. 
According to Fig.~\ref{fig:timemem}, 
$\alpha^\alpha$-Rank can achieve three orders of magnitude reduction in time (i.e. $1000x$ faster) compared to default $\alpha$-Rank implementation from  \cite{lanctot2019openspiel}.
Memory-wise, our method uses only half of the space when considering, for instance, matrices of the size $10^4 \times 10^4$.

\textbf{{{Autonomous Driving on Highway:}}} Having assessed correctness and scalability, we now present novel application domains on large-scale \emph{multi-agent/multi-player} systems. For that we made used of high-way \cite{highway-env}; an environment for simulating  self-driving scenarios with social vehicles designed to mimic real-world traffic flow.
We  conducted a ranking experiment involving $5$ agents each with $5$ strategies, i.e., a strategy space in the order of $\mathcal{O}(5^{5})$ ($3125$ possible strategy profiles). 
Agent strategies varied between ``rational'' and ``dangerous'' drivers, which we encoded using different reward functions during training (complete details of reward functions are in \href{https://drive.google.com/file/d/1xz3OMSM8uUc3DpHCke-gqXq_Z37cb_ea/view}{Appendix}).
Under this setting, we knew, upfront, that optimal profile corresponds to all agents being five rational drivers. Cars trained using value iteration and the rewards averaged from 200 test trails were reported. 

We considered both $\alpha^\alpha$-Rank and $\alpha^\alpha$-Oracle, and reported the results by running  $1000$ random seeds.
We set $\alpha^\alpha$-Oracle to run $200$ iterations of gradient updates in solving the top-rank strategy profile (lines $8-12$ in Algorithm~\ref{alpharank_alg}). 
Results depicted in Fig.~\ref{fig:ising_car}(a) clearly demonstrate that both our proposed methods are capable of recovering the correct highest ranking strategy profile. 
$\alpha^\alpha$-Oracle converges faster than $\alpha^\alpha$-Rank, which we believe is due to the oracle mechanism saving time in inefficiently exploring ``dangerous" drivers upon one observation.  
We also note that although such size of problem are feasible using $\alpha$-Rank and the Power Method, our results achieve 4 orders of reduction in number of iterations. 

\textbf{{{Ising Model Experiment:}}} We repeated the above experiment on the Ising model \cite{ising1925beitrag} that is typically used for describing ferromagnetism in statistical mechanics. It assumes a system of magnetic spins, where each spin $a^j$ is either an up-spin, $\uparrow$, or down-spin,       $\downarrow$.
The  system energy is defined by {\small $E(\textbf{a}, h)=-\sum_{j}(h^{j} a^{j}+\frac{\lambda}{2} \sum_{k \neq j} a^{j} a^{k})$} with $h^j$ and $\lambda$ being constant coefficients. 
The probability of one spin configuration is {\small $P(\textbf{a})=\frac{\exp (-E(\textbf{a}, h) / \tau)}{\sum_{a} \exp (-E(\textbf{a}, h) / \tau)}$ } where $\tau$ is the environmental temperature.
Finding the equilibrium of the system is notoriously hard because it is needed to enumerate all possible configurations  in computing $P(\textbf{a})$.
Traditional approaches include Markov Chain Monte Carlo (MCMC). 
An interesting phenomenon  is the \emph{phase change}, i.e., the spins will reach an equilibrium in the low temperatures, with the increasing $\tau$, such equilibrium will suddenly break and the system becomes chaotic. 

Here we try to observe the phase change through multi-agent evaluation methods. 
We assume each spins as an agent, and  the reward  to be $r^{j}=h^{j} a^{j}+\frac{\lambda}{2} \sum_{k \neq j} a^{j} a^{k}$. We consider the top-rank strategy profile from $\alpha^\alpha$-Oracle as the system equilibrium      
and  compare it against the ground truth from MCMC.
We consider a $5\times5$ 2D model which induces a prohibitively-large strategy space of the size $\boldsymbol{2^{25}}$ ($\approx \text{$33$ million strategies}$) to which existing methods are inapplicable.
Fig.~\ref{fig:ising_car}(b) illustrates that our method  identifies  the same phase change as that of MCMC. We also show an example of how $\alpha^\alpha$-Oracle's top-ranked profile finds the system's equilibrium  when $\tau=1$ in Fig.~\ref{fig:ising_car}(c). 
Note that the  problem of $25$ agent with $2$ strategies goes far beyond the capability of $\alpha$-Rank  on one single machine (billions of elements in $\bm{T}$); we therefore don't list its performance here.

\section{Conclusions \& Future Work}
In this paper, we presented major bottlenecks prohibiting $\alpha$-Rank from scaling beyond tens of agents. 
Dependent on the type of input, $\alpha$-Rank's time and memory complexities can easily become exponential. We further argued that notions introduced in $\alpha$-Rank can lead to confusing tractability results on notoriously difficult
 NP-hard problems. To eradicate any doubts, we empirically validated our claims by presenting dollars spent as a non-refutable metric. 

Realising these problems, we proposed a scalable alternative for multi-agent evaluation based on stochastic optimisation and double oracles, along with rigorous scalability results on a variety of benchmarks. 
 For future work, we plan to understand the relation between $\alpha$-Rank's solution and that of a Nash equilibrium. Second, we will attempt to conduct a theoretical study on the convergence of our proposed $\alpha^{\alpha}$-Oracle algorithm.
\clearpage

\bibliographystyle{ACM-Reference-Format}  
\bibliography{iclr2020_conference.bib}

\end{document}